\documentclass{emulateapj}

\usepackage{amssymb}
\usepackage{amsmath}

\def\aj{\rm{AJ}}                    
\def\apj{\rm{ApJ}}                 
\def\apjl{\rm{ApJ}}                
\def\apjs{\rm{ApJS}}                       
\def\mnras{\rm{MNRAS}}   

\usepackage{epsfig}
\shorttitle{The Impossibly Early Galaxy Problem}
\shortauthors{Steinhardt et al.}
\begin{document}
\title{The Impossibly Early Galaxy Problem}
\author{Charles. L. Steinhardt \altaffilmark{1,2}, Peter Capak\altaffilmark{1,2}, Dan Masters\altaffilmark{1,2}, Josh S. Speagle\altaffilmark{3,2,4}}

\altaffiltext{1}{California Institute of Technology, MC 105-24, 1200 East California Blvd., Pasadena, CA 91125, USA}
\altaffiltext{2}{Infrared Processing and Analysis Center, California Institute of Technology, MC 100-22, 770 South Wilson Ave., Pasadena, CA 91125, USA}
\altaffiltext{3}{Harvard University Department of Astronomy, 60 Garden St., MS 46, Cambridge, MA 02138, USA}
\altaffiltext{4}{Kavli Institute for the Physics and Mathematics of the Universe, Kashiwanoha 5-1-5, Kashiwa-shi, Chiba-ken, 277-0882, Japan}

\begin{abstract}
The current hierarchical merging paradigm and $\Lambda$CDM predict that the $z \sim 4-8$ universe should be a time in which the most massive galaxies are transitioning from their initial halo assembly to the later baryonic evolution seen in star-forming galaxies and quasars.  However, no evidence of this transition has been found in many high redshift galaxy surveys including CFHTLS, CANDELS and SPLASH, the first studies to probe the high-mass end at these redshifts.  Indeed, if halo mass to stellar mass ratios estimated at lower-redshift continue to $z \sim 6-8$, CANDELS and SPLASH report several orders of magnitude more $M \sim 10^{12-13} M_\odot$ halos than are possible to have formed by those redshifts, implying these massive galaxies formed impossibly early.  We consider various systematics in the stellar synthesis models used to estimate physical parameters and possible galaxy formation scenarios in an effort to reconcile observation with theory.  Although known uncertainties can greatly reduce the disparity between recent observations and cold dark matter merger simulations, even taking the most conservative view of the observations, there remains considerable tension with current theory. 
\end{abstract}



\section{Introduction}

Current theory predicts that galaxies begin their existence as tiny density fluctuations, with overdense regions collapsing into virialized protogalaxies and eventually assemble gas and dust into stars and black holes.  Although the details of these later stages of assembly continue to present a difficult theoretical problem, there is broad consensus on the mass and redshift distribution of halos produced through an initial collapse and hierarchical merging \citep{Sheth2001,Springel2005,Vogelsberger2014}.  For any specific model, including the consensus hierarchical merging model with $w = -1$ dark energy and without exotic dark matter, the halo mass function is straightforward to calculate.  The predicted halo mass function is also potentially a sensitive probe of the effects of dark matter and dark energy, with warm dark matter models inhibiting halo formation \citep{Lovell2014} and $w > -1$ dark energy models allowing earlier massive halo formation \citep{OHara2006,Gladders2007}.  The consensus in these models is a rapid evolution in the density of massive haloes at $z>4$ that should be observationally evident in galaxy luminosity and mass functions. 

Until recently, observations of the highest-redshift galaxies were limited to a small number of massive individual galaxies and quasars at $z > 6$ (e.g., \citet{Fontana2006,Mortlock2011}).  The recent advent of high-redshift surveys including the Cosmic Assembly Near-infrared Deep Extragalactic Survey (CANDELS; \citet{CANDELS}) and the Spitzer Large Area Survey with Hyper-Suprime-Cam (SPLASH; Capak et al. in prep.) now allows us to probe the galaxy luminosity and mass functions, and by implication the corresponding halo mass function, across a range of masses from $z=4-8$.  In particular, SPLASH has the broad sky coverage required to measure the number density of relatively rare, massive galaxies, while CANDELS is a deeper survey over a smaller area, more suited to finding lower-mass galaxies.

A range of work has estimated halo masses for $z>4$ galaxies and pointed out the tension between the expected evolution of the halo mass function and galaxy luminosity and mass function (e.g., \cite{Lee2012}, \citet{Hildebrandt2009}, \citet{Finkelstein2015}, \citet{Caputi2015}).  Here we assemble this previous work and update it with more recent measurements of the galaxy luminosity function, our knowledge of the $z>4$ galaxy population, and stellar synthesis models.  We begin by showing that the rapid evolution of the high-mass end of the halo mass function at $z \sim 4-8$ allows the rare, most luminous and/or massive galaxies in large surveys to provide stringent constraints on galaxy evolution in the hierarchical merging paradigm.  The most straightforward prediction comes from the mass and number density of the most massive halos that have formed as a function of redshift.  Furthermore, the consensus $\Lambda$CDM model combined with our understanding of galaxy evolution also predicts a sharp drop in the number density of high-luminosity galaxies at fixed $L$ and a rapid decline in the luminosity of massive galaxies at fixed number density towards high redshift.  The details of this rapid evolution, if observed, will be sensitive to the baryonic physics of early star formation and $\Lambda$CDM models.  

In \S~\ref{sec:highm}, we show that the most luminous high-redshift galaxies apparently lie in halos too massive to have been able to have formed via hierarchical merging by those redshifts.  In \S~\ref{subsec:normal} we show the physical properties of high-mass galaxies lie on several scaling relations derived at lower redshift, so that it is natural to believe they have been inferred correctly.  However, because this result relies heavily upon assumptions about the halo mass to light ratio, we are motivated to develop alternative tests of hierarchical merging based directly on luminosity functions, that allow us to control for systematic errors.  Using these tests, in \S~\ref{subsec:lum}, we show that high-luminosity galaxies do not exhibit the rapid evolution in number density at fixed $L$ expected from hierarchical merging.  We also discuss the expected conversion between luminosity and halo mass, showing that an implausibly sharp evolution in $M_{Halo}/L$ would be required for consistency with theoretical predictions.  Several ways of altering galaxy formation models in order to produce these ``impossibly early galaxies" are considered in \S~\ref{sec:models}.  We consider the implications of these disagreements in \S~\ref{sec:discussion}.

This work uses a $(h, \Omega_m, \Omega_\Lambda) = (0.704, 0.272, 0.728)$ cosmology \citep{Planck2015} throughout.  Unless otherwise specified, mass-to-light ratios refer to rest-frame UV wavelengths throughout.

\section{The Earliest Galaxies and their Halos}
\label{sec:highm}

In the consensus $\Lambda$CDM model, the high-mass end of the predicted halo mass function changes rapidly between $z \sim 8-4$, with halos containing the most massive galaxies typically virializing towards $z = 4$ (e.g., \citet{Sheth2001}).  The timespan of 0.9 Gyr over this redshift range means that we likely observe these galaxies within at most a few dynamical times of their initial assembly.  Since galaxies are expected to form after their halos assemble, the number density of massive systems and its redshift evolution can provide a good probe of the initial formation of the their dark matter halos.  The broad redshift range over a relatively small amount of time allows for more precise cosmic epoch measurements than are easily obtainable at lower redshift.  

However, at these high redshifts there are a limited number of tools available for estimating halo masses.  The most robust are galaxy clustering based methods that infer the halo masses from the spatial distribution of galaxies (e.g., \citet{Hildebrandt2009,Lee2012}).  Clustering methods have the advantage of not requiring assumptions about the physical properties of the galaxies, but must assume a model for the dark matter from simulations.  

Other methods rely on a relationship between luminosity or stellar masses estimated from template fitting (cf. \citet{Ilbert2013}) to infer halo masses. ``Abundance matching"  (e.g., \citet{Finkelstein2015}) ties a key feature of a luminosity or mass function, such as the knee of the luminosity function, to a feature halo mass function, then matches the galaxy density and the dark matter halo density to derive halo masses.  The main shortcoming of this method is that it will reproduce the halo mass function by definition and is sensitive to the point at which the two functions are tied together.  Finally, one can assume a luminosity or stellar mass to dark matter mass ratio measured at lower redshifts where better estimators of dark matter are available \citep{Leathaud2012}.

In the past, only a handful of $z>4$, galaxies were known, far too few to estimate the corresponding halo mass function using any of these methods.  However, recent infrared observations by the CANDELS collaboration \citep{Duncan2014,Grazian2015} and the Spitzer Large Area Survey with Hyper-Suprime-Cam (SPLASH; \citet{Steinhardt2014a}) provide a much larger sample of high-redshift galaxies with stellar mass estimates and can now be used to estimate the halo mass function at $z \sim 6-8$.  

\begin{figure}
\plotone{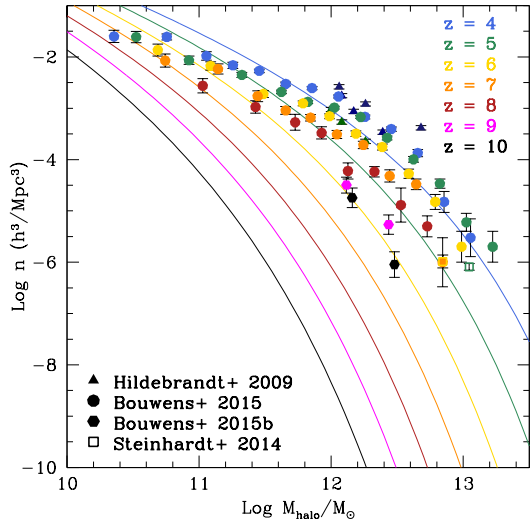}
\caption{Theoretical halo number density as a function of halo mass and redshift \citep{Sheth2001,HMFcalc} for the most massive halos at $4 < z < 10$ (shown as solid lines, with redder colors at higher redshift) compared with observational number densities of estimated halo masses corresponding to observed star-forming galaxies at similar redshifts.  Halo masses are estimated using clustering (triangle), stellar masses converted to halo masses using thelow-redshift scaling ratio $M_H/M_* \sim 70$ (square), and UV luminosities converted to halo masses using ratios determined by lower-redshift abundance matching (circle), as described in \S~\ref{sec:highm}, for an overall $M_H/M_\odot \approx 120 L_{UV}/L_\odot$.  These methods all give self-consistent number densities that disagree with theoretical expectations.  We discuss these methods and possible sources of error using these techniques in more detail in \S~\ref{subsec:normal}.}
\label{fig:z6data}
\end{figure}

A subset of these halo mass function measurements are compiled in Figure \ref{fig:z6data}, using three different techniques depending upon the type of observation used.  For points derived a clustering analysis (triangles), the halo mass function is measured as a direct result of the technique, and we report the derived halo mass and number density.  For points derived using photometric SED template fitting, the best-fit inferred stellar mass function is convert to halo mass using $M_H/M_* \sim 70$ derived at lower redshift (cf. \citet{Leathaud2012}).  Finally, for UV luminosity functions, the monochromatic UV luminosity is converted to stellar mass using the mean stellar mass-to-light ratio derived from abundance matching at $z=4$, where clustering analysis is also available, then to a halo mass using $M_H/M_* \sim 70$.  The overall conversion is $M_H/M_\odot \approx 120 L_{UV}/L_\odot$.

A subset of these halo mass function measurements are compiled in Figure \ref{fig:z6data}, showing the large and diverging disagreement between the theoretical and observational evolution of the halo mass function at high redshift.  We compare them with theoretical halo mass functions estimated using \citet{Sheth2001}, as provided by HMFCalc \citep{HMFcalc}.  Specifically, we find that observational halo mass function estimates correspond to a higher number density of massive halos than should have been able to form through the rapid collapse and evolution of rare, highly-overdense regions (Fig. \ref{fig:z6data}).  

One possible explanation is that the $M_H/L_{UV}$ ratio in massive galaxies sharply decreases at $z>4$, leading to overestimated halo masses for high-redshift galaxies.  If so, we might hope that this rapid evolution should be evident from other measured properties of the galaxy population.  In \S~\ref{subsec:normal}, we will use scaling relations to probe this possibility, instead finding that this rapid evolution is not observed.

\section{High-Redshift Galaxies Appear `Normal'}
\label{subsec:normal}
We have shown that there appears to be a sharp conflict between the halo mass function inferred from observations at $z \sim 4-8$ and the theoretical production of halos in the early universe.  We term this the {\em impossibly early} galaxy problem.  However, many of these estimates critically depend on galaxy spectral energy distribution templates and scaling relations derived at lower redshift.  As a result, the possibility that these assumptions have broken down by $z \sim 4-8$ is a major concern (cf. \citet{Conroy2009,Schaerer2010,Lee2012,Stark2013,Speagle2014}), and such a breakdown would provide an easy solution to the impossibly early galaxy problem.  

Lower-redshift galaxies are much more robustly understood, and follow several trends and scaling relations, both at fixed redshift and in their redshift evolution.  A good test of whether high-redshift galaxies are truly similar to low-redshift galaxies, as has been strongly assumed, is to check whether they lie on the high-redshift extrapolation of this behavior.  As summarized below, we indeed find that early, high-redshift galaxies appear to be `normal', including the most massive early galaxies that provide the sharpest disagreement with theoretical halo production.

Over the past few decades, there have been a series of observational results indicating that many processes in the most massive and luminous galaxies, including star formation \citep{Cowie1996,Madau1998,PerezGonzalez2008} and quasar accretion \citep{Richards2006a,Labita2009,Steinhardt2010c}, occur earlier than in less massive galaxies.  The early evolution of star forming galaxies is further supported by the existence of a population of massive, passive galaxies at $z \sim 2-4$ \citep{Glazebrook2004,Daddi2005,Carollo2013,Ilbert2013,Straatman2014}, which must have been massive star-forming galaxies at high redshift similar to those reported by SPLASH and CANDELS \citep{Toft2012}.  Furthermore, Fe{\small II}/Mg{\small II} ratios in high-redshift quasars \citep{Barth2003,Kurk2007} and ages derived from template fitting \citep{Maraston2010} suggest that early, massive galaxies indeed form their stars very rapidly.  

In contrast, $\Lambda$CDM theory requires that more massive haloes generally assemble later than less massive ones. So, reconciling the theory and observations requires that more massive galaxies must evolve far more rapidly than less massive ones.  Because the dynamical time increases with mass, this rapid evolution would require a change in the baryonic processes that dominate star formation at lower redshifts.  Such a model would predict that the redshift $z \sim 4-8$ universe is a period of rapid transition between the initial assembly of massive halos and the quick growth of their stellar populations.   

\subsection{Similar SFR-Stellar Mass Relations}
To date, SPLASH and CANDELS find no apparent deviation from properties derived at lower redshift. Instead, they find that the trend towards more and more massive galaxies evolving earlier continues out to $z \sim 6-8$.  In particular, at $0 < z < 4$ it has been shown that star-forming galaxies lie on a ``main sequence'', with a tight correlation between the existing stellar mass and their star-formation rate (cf. \citet{Peng2010}, \citet{Noeske2007}).  An analysis of over two dozen studies of star-forming galaxies using different techniques for selection, for measuring stellar mass, and for measuring star formation rate shows strong agreement in both slope and redshift evolution \citep{Speagle2014}.  Early, high-redshift galaxies lie on the extrapolation of this main sequence to $z \sim 6-7$.

The star-forming main sequence is well-measured for lower-mass objects at $z \sim 5-6$, with agreement between SPLASH \citep{Steinhardt2014a} and CANDELS \citep{Duncan2014}.  Where stellar masses are available, the most massive early galaxies lie directly on the high-mass extrapolation of this main sequence.  Because the stellar mass and star formation rates are measured using different wavelengths, systematic errors from incorrect template fitting would produce incorrect inferred quantities in different ways, and likely produce a population inconsistent with the main sequence.  Similarly, arbitrary deblending problems would produce outliers lying somewhere arbitrary, and thus likely off this main sequence. 

\begin{figure}
\plotone{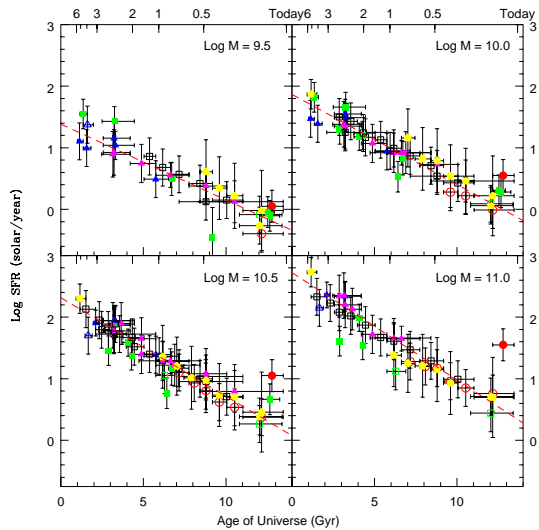}
\caption{96 measurements from 32 studies of the star-forming main sequence\altaffilmark{4} for fixed $M_* = 10^{10.5}M_\odot$ at $0 < z < 6$, adjusted to lie on a common set of calibrations following the prescription derived by \citet{Speagle2014} using 25 of these studies.  The horizontal bars indicate the redshift range spanned by each particular observation, while the vertical errors are the “true” scatter about each MS observation \citep{Speagle2014}.  Although these studies use many different methods for determining SFR and stellar mass (blue = UV, purple = UV+IR, red = IR, green = emission lines, yellow = SED fitting, black = radio), they all show good agreement with a best-fit (log) linear evolution determined at $0 < z < 4$ (dotted line), and the highest-redshift measurements lie are consistent with the extrapolation of that fit out to $z \sim 6$.  It therefore seems likely that techniques for estimating stellar masses and star formation rates have not become catastrophically incorrect at $z \sim 6$.}
\label{fig:speagle2014}
\end{figure}

The redshift evolution of the star-forming main sequence is also well understood at $0 < z < 4$, and the observed $4 < z < 6$ star-forming main sequence has the properties produced by extrapolating that time-dependence.  The increase in mass towards high redshift of the $N$ most massive star-forming galaxies (one component of what has been termed the "downsizing" problem) is also well measured \citep{vanDokkum2010,Duncan2014,Steinhardt2014c}, and the most massive star-forming galaxies at $z \sim 6$ lie on the extrapolation of the best-fit time-dependence at $0 < z < 4$ (Fig. \ref{fig:speagle2014}).

At some point stellar templates derived from low redshift will cease to be valid, but this has yet to be observed.  It might be expected that this occurs only when the physics of star formation have changed, perhaps due to very low metallicities producing a very top-heavy IMF and even Population III stars.  If so, templates may continue to be valid well above $z \sim 6-8$.  At a minimum, it seems clear that they have not become catastrophically incorrect in the redshift range where impossibly early galaxies have been measured.

\subsection{Quasar-Host Galaxy Relations}
A final test of consistency for early, massive star-forming galaxies comes from quasar accretion.  It is believed that there is coevolution between a galaxy and its central supermassive black hole.  Quasar accretion exhibits similar behavior to star-forming galaxies in the sense that there is ``main sequence''-like behavior \citep{Richards2006a,Labita2009,Steinhardt2010c,Shen2011} with the most luminous quasars becoming more luminous and having larger black hole masses towards higher redshift.  Indeed, there has historically been an ``impossibly early black hole'' problem similar to the impossibly early galaxies presented in this work (\citet{Mortlock2011}; cf. \citet{Carr2003}, \citet{Madau2001}, \citet{Bromm2003}), although as we discuss in this work the star formation problem appears to be more difficult to solve.  

\footnotetext{25 of these studies were included in \citet{Speagle2014}, with \citet{Magnelli2014}, \citet{Heinis2014}, \citet{Whitaker2014}, \citet{Duncan2014}, \citet{Schreiber2015}, \citet{Pannella2015}, and \citet{Salmon2015} added for this figure.}

One parameterization of the similarity between the redshift evolution of quasars and star-forming galaxies is to note that the ratio of $M_*$ for most massive star-forming galaxies to $M_{BH}$ for the most massive quasars is observed to be approximately 30:1 at all fixed redshifts (as shown by \citet{Steinhardt2014c} from a literature compilation).  The $\log M_*/M_\odot \sim 11.2$ galaxies at $z \sim 6$ in SPLASH have a similar ratio with early, massive quasars such as the $\log M_{BH}/M_\odot \sim 10.08$ quasar recently reported at $z = 6.4$ \citep{Wu2015}.  Because virial black hole mass estimates have very different systematic uncertainties than stellar mass estimates, this ratio continuing to hold is additional evidence that the properties inferred for impossibly early galaxies are likely reasonable.

Certainly at some sufficiently high redshift, these various scaling relations and correlations should break down, but again this has yet to be observed.  The key point is that the observed high-redshift population of star-forming galaxies follows the expected redshift evolution determined observationally from many lower-redshift surveys.  Even the most massive, highest-redshift objects are found and analyzed using the same standard techniques that have been verified to be successful at lower redshift, and are remarkable solely because of the discrepancy between observation and $\Lambda$CDM, not because of the data.  

\section{Luminosity Functions as Probes of Halo Mass}
\label{subsec:lum}

UV luminosity functions are one of the cheapest and most robust observations to obtain at high redshift, and can be converted to halo masses by assuming a halo mass-to-light ratio.  Indeed, most of the high-redshift data shown in Fig. \ref{fig:z6data} and all of those at $z > 6$ are derived from monochromatic UV luminosity functions \citep{McLure2009,Lee2012,Bouwens2015,Bouwens2016}, the only observations available.  The main advantage of this technique over mass functions is that it places most of the potentially large uncertainty into the halo mass-to-light ratio, which can be parameterized in a single model where it can be analyzed more clearly.  In contrast, mass functions place much of the uncertainty in the details of the diverse set of spectral energy distribution models fit to each individual galaxy where they can be difficult to understand \citep{Conroy2009}.

\begin{figure}
\plotone{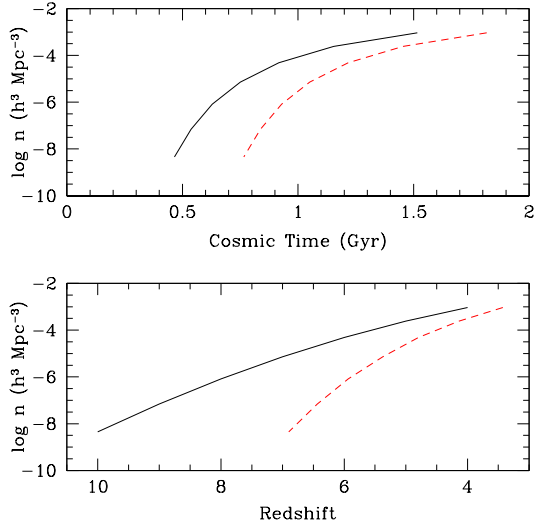}
\caption{Number density of $10^{12} M_\odot$ halos (black, solid) from theoretical predictions and corresponding number density (i.e., point on a luminosity function) for star-forming galaxies (red, dashed) for a toy model in which the stars form 300 Myr after halo virializtion.  There is a characteristic time for forming such halos, seen as a sharp rise in number density over time (top) or towards higher redshift (bottom).  Although the rise in number density of halos and (300 Myr later) galaxies is equally fast in time, because of the age-redshift relation, the rise in galaxies, taking place at a later time and lower redshift, appears sharper with respect to redshift.  Because the evolution of mass and luminosity functions is typically shown in terms of redshift evolution, this means that a sharp rise in number density of halos appears to produce an even sharper obse rved evolution in the corresponding luminosity function.}
\label{fig:zt}
\end{figure}

In addition to uncertainty in the correspondance between halo mass and UV luminosity, comparing observation with theory also requires matching observations of galaxies in the midst of star formation with theoretical predictions about the time at which their halo was formed.  After all, galaxies likely do not emerge fully-formed at the moment a halo virializes; the dark matter merely has to collapse gravitationally, while baryons get a later start at small masses (cf. \citet{Haiman1997}) and must clump and cool in order to form stars.  If this process typically takes, e.g., 300 Myr for every galaxy due to dynamical considerations, then a sharp rise in the halo mass function as predicted from redshift 8.0 to 4.0 would lead to a sharp rise in the UV luminosity function for corresponding galaxies from redshift 6.0 to 3.4.  More generally, in a matter-dominated universe, $dz/dt \propto t^{-5/2}$.  Thus, delays between halo (smaller $t$) and galaxy formation (larger $t$) lead to the UV luminosity function evolving equally rapidly with time but {\it over a smaller range in redshift}, producing a more noticeable effect observationally (Fig. \ref{fig:zt}).

These two uncertainties -- how to match star-forming galaxy UV luminosities to halo formation in both mass and time -- both lie at the heart of the key open questions concerning early galactic evolution.  Determining these two parameters requires a better understanding of several key aspects of galactic evolution.  The time delay between assembly and later evolution depends upon the relationship between halo formation and the evolution of the galaxies that evolve inside those halos.  Constraining the luminosity to halo mass ratio involves constraining the high-redshift initial mass function, dust content, and other processes involved in the formation of the first large stellar populations made by galaxies.  

Because there is very little degeneracy between shifts in mass and time, if both the UV luminosity function and halo mass function are well measured independently, it should be possible to determine both shifts.  Indeed, such a determination is the core idea behind abundance matching, using a combination of theoretical predictions and measurements of number density to `match' halos with galaxies (e.g., \citet{Lee2009,Behroozi2015,Finkelstein2015}.  What we find, however, is that this matching process seems to fail at high redshift.  As shown in Fig. \ref{fig:z6data}, matching halos with galaxies would require that galaxies live in halos so small as to be unphysical, in extreme cases less than a factor of three higher than stellar masses.  Further, the sharp high-mass evolution in the halo mass function has no observed counterpart in luminosity functions, either at high redshift or over a similar period of time at lower redshift.

\subsection{Halo Mass to Light Ratio Evolution}
\label{subsec:mlevolve}

Out to $z = 4.7$, several studies \citep{Hildebrandt2009,Lee2012,Finkelstein2015,McLure2009,Ouchi2009}) are consistent with a nearly constant halo mass-to-light ratio.  However, this ratio must evolve at higher redshifts to reconcile the UV luminosity function with hierarchical merging (Fig. \ref{fig:z6data}).  There are two ways this might occur: halos might be more efficient at producing stars than previously believed or stars might produce more UV light per unit stellar mass than at lower redshift.  

In order to determine the best explanation, consider the size of the change in the halo mass to luminosity ratio that would be required to reconcile observation with theory.  At a fixed monochromatic luminosity of, for example, $M_{1600,AB} = -21$, there is a 1.5 dex decrease in number density from $z=4$ to $8$ \citep{Bouwens2015}.  At $z \sim 4$, clustering analysis shows that this corresponds to a halo mass of $\log M_{Halo}/M_\odot = 12.4$ \citep{Hildebrandt2009}.  In the most (very likely overly) optimistic scenario, in which stars instantly come into existence so that there is no time delay between halo and galaxy formation, the observed number density for $M_{1600,AB} = -21$ galaxies at $z = 8$ corresponds to the number density of $\log M_{Halo}/M_\odot = 11.6$ halos.  Thus, a sharp evolution of 0.8 dex in $M_{Halo}/L_{UV}$ would be required from $z = 4$ to $8$.  We consider in the remainder of this section whether this evolution is plausible.

\subsubsection{Stellar Evolution Models}
\label{subsec:asp}

Evolution in the halo mass to light ratio could be considered to be dominated by four effects:
\begin{enumerate}
\item{{\bf Stellar Evolution:} In a typical stellar population, the luminosity is dominated by massive stars, but the stellar mass (which is used to estimate the halo mass) is dominated by low-mass stars.  As a result, the mass to light ratio is higher for older stellar populations.  Assuming galaxies at $z = 8$ have younger stellar populations than $z = 4$, this effect acts in the direction of reconciling observation with theory.  Increasing metallicity associated with maturing stellar populations enhances this effect.}
\item{{\bf Changing IMF:} If the initial mass function (IMF) is top-heavy at high redshift, as expected for Pop. III stars, this would again act to decrease the mass to light ratio.  However, the IMF is expected to be similar at all $z < 8$ based upon seemingly analagous low-redshift systems \citep{Dias2010}, and therefore to have no impact on impossibly early galaxies (\S~\ref{subsubsec:imf}).}
\item{{\bf Evolving Dust Corrections:} Very few galaxies with photometric $z > 4$ are selected as highly dusty (cf. \citet{Ilbert2010}), so a strong reduction in extinction from current models is not plausible.  However, if they were dusty, an increase in extinction would increase the halo mass to monochromatic UV luminosity ratio, making the halos containing early galaxies even more massive than currently measured.}
\item{{\bf Merging and Time Delays:} Clustering and merging results in the addition of both halo mass and stellar mass to galaxies, as well as additional gas that will eventually form stars.  A merger of two large objects with the same mass-to-light ratio will yield a galaxy with that same ratio.  However, more gradual accretion of both dark matter and gas capable of forming stars will produce an immediate increase in halo mass but delayed increase in stellar mass, making the typical halo mass-to-light ratio larger at $z = 8$ than at $z = 4$.  This would act in the direction of making the problem worse, because galaxies at the same luminosity would now reside in even more massive halos, which have a lower number density and later virialization time.  For example, \citet{Behroozi2015} find that a $10^{10} M_\odot$ galaxy at $z=7$ should correspond to a $10^{12.9} M_\odot$ halo, whereas we have assumed based upon lower-redshift measurements that its halo was below $10^{12} M_\odot$.}

\end{enumerate}

\begin{figure}
\plotone{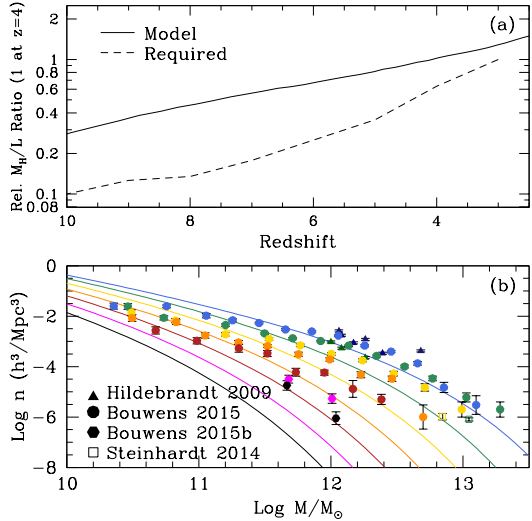}
\caption{(a) Expected halo mass to monochromatic UV luminosity ratio, along with the required evolution to reconcile observation with theory, and (b) resulting corrected halo mass functions derived as in Fig. \ref{fig:z6data} with $M_{halo}/L_{UV}$ evolving due to a stellar population starting at low metallicity at $z = 12$ and aging along the star-forming main sequence, as described in \S~\ref{subsec:asp}.  Such a model would be reasonable given observational constraints, but cannot produce agreement between measured UV luminosity functions and simulated halo mass functions.}
\label{fig:realistic}
\end{figure}

The best-case realistic scenario for producing early, luminous galaxies, then, is one in which an isolated stellar population is aging as rapidly as possible (\#1) and there is no evolution in dust (\#3) or time delay between infall and star formation (\#4).  We model the halo mass to light ratio (Fig. \ref{fig:realistic}a) from an initial stellar population that formed in one rapid burst, perhaps as early as $z = 12$ \citep{Planck2015}, followed by evolution along the star-forming main sequence until observed at $z = 4-8$.  If the time dependence of the star-forming main sequence at $z > 6$ can be extrapolated from the best-fit evolution at $z < 6$ (as determined by, e.g., \citet{Speagle2014}), it will be nearly constant at higher redshifts because of the short gap in elapsed time.  Thus, SFR $\propto M_*^{0.7}$, with the stellar age increasing from an initial small value to asymptotically approach 50-150 Myr. depending upon redshift.  This is insufficient to reconcile observation with theory (Fig. \ref{fig:realistic}b).  

\begin{figure}
\plotone{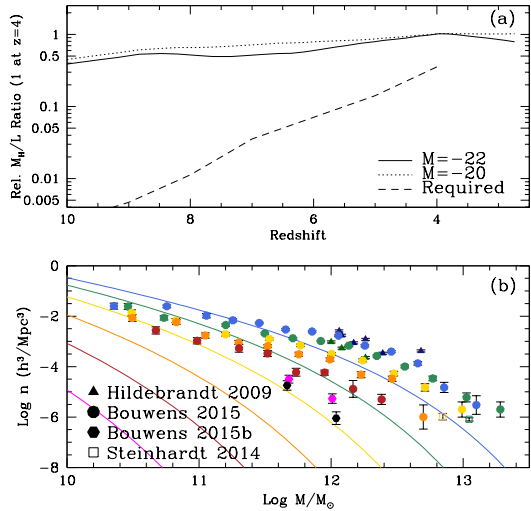}
\caption{Halo mass and UV luminosity functions for a population evolving due to (1) a stellar population starting at low metallicity at $z = 12$ and aging along the star-forming main sequence, as described in \S~\ref{subsec:asp}; (2) observed evolution in mean extinction as a function of redshift and luminosity; and (3) a 300 Myr time delay between dark matter accretion and star formation within halos, so that observed UV luminosity functions are compared with merger simulation results at higher redshifts.  (a) Expected halo mass to monochromatic UV luminosity ratio at two different absolute magnitudes along with the required evolution in the halo mass to luminosity ratio to reconcile observation with theory.  Observed dust corrections are magnitude-dependent, and therefore so is the mass-to-light ratio.  (b) Resulting corrected observational halo mass functions derived as in Fig. \ref{fig:z6data}, with theoretical halo mass functions additional shifted by a 300 Myr time delay, so that the yellow $z=6.00$ ($t=0.954$ Gyr) observationally derived halo masses are matched with a yellow $z=8.01$ ($t=0.654$ Gyr) halo mass function.  Such a model fits well with current observational constraints, but cannot reconcile observation with theory.}
\label{fig:morerealistic}
\end{figure}

The most likely scenarios are even more difficult to reconcile with theory.  We model the evolution of the halo mass to light ratio (Fig. \ref{fig:morerealistic}a) as above, but further include a 300 Myr delay (cf. \citet{Wong2009}) between merging and resulting star formation motivated by dynamical timescales, so that the luminosity function at time $t$ is used to predict the halo mass function at $t - 300$ Myr.   The halo mass functions in Fig. \ref{fig:morerealistic}b are colored to match the redshift of the corresponding luminosity function, so that the yellow $z=6.00$ ($t=0.954$ Gyr) observationally-derived halo masses are matched with a yellow $z=8.01$ ($t=0.654$ Gyr) halo mass function.

We additionally include dust evolution as in \citet{Bouwens2015} from $z = 8$ to $z = 4$.  This has no effect on halo mass estimates derived from clustering and from template fitting that already includes extinction, but will alter halo masses estimated from UV luminosities.  For the purposes of this model, we take the $z = 4$ abundance matching-derived halo mass to UV luminosity ratios as correct, then apply an additional correction at higher redshift to account for evolution in the mean extinction at that luminosity between that redshift and $z = 4$.  The resulting comparison between observation and theory (Fig. \ref{fig:morerealistic}b) is likely our best model given current data, and produces a larger disagreement than the models previously discussed.

Perhaps the most realistic idea along these lines would be to create all of the stars directly at $z=8$, in which case the age-zero stellar population decreases the halo mass-to-light ratio sufficiently to produce a match between the $z=8$ halo mass and UV luminosity function.  However, this would require these galaxies by $z = 4$ to have stellar populations nearly 1 Gyr old (and thus appear passive), and there should be a significant population of high-redshift starbursting galaxies. Instead, massive galaxies at these redshifts are almost exclusively star-forming \citep{Ilbert2010,Lee2012,Ilbert2013,Stefanon2013,Steinhardt2014a}, typically with $<100$ Myr old stellar populations.  Furthermore, there is an apparent drop in the density of extreme starburst galaxies towards higher redshift \citep{Casey2012}, contradicting the required evolution.  

Even discarding these apparently contradictory observations, this sharp shift is not quite enough to reconcile observed luminosities with halo mass functions because these stellar populations must then age.  It is possible in this unphysical model to drive the inferred stellar masses low enough to avoid a problem directly at the redshift when all of the star formation is assumed to take place ($z = 8$ in this example), but it will reappear a short time later and is as sharp of a conflict as before by $z = 6$.

Another approach is to allow an inconsistency between halo masses and inferred baryon masses in order to match halo mass with light.  
\citet{Finkelstein2015} used mass-to-light ratios to determine $M_*$ and abundance matching to determine $M_{Halo}$ from CANDELS, which requires that $M_{Halo}/M_{baryon}$ appears to evolve from $\sim 40:1$ at $z \sim 8$ to $\sim 100:1$ by $z \sim 4$, similar to the overall ratio of 70:1 by low redshift \citep{Planck2015}.  In the absence of exotic high-energy physics, dark matter would collapse at least as early as baryons.  Therefore, it is difficult to explain how this ratio can increase over time.  

However, a stellar population aging due to evolution along the star-forming main sequence provides an alternative explanation.  Consider a low-metallicity galaxy beginning with a rapid starburst at $z = 10$ followed by main sequence star formation.  As the galaxy grows, the average age of the stellar population increases as well, and $M_*/L_{UV}$ increases from $z = 8$ to $z = 4$.  Using the $z = 4$ value of $M_*/L_{UV}$ at all redshifts will correctly estimate $M_*$ at $z = 4$, but overestimate $M_*$ by a factor of 2--3 at $z  = 8$.  Assuming that $M_{baryon}/M_*$ remains constant, correcting the stellar masses would produce $M_{Halo}/M_{baryon} \sim 100:1$ at $z = 8$.  Thus, a reasonable evolution in the age of stellar populations could reconcile abundance matching with our expectation of constant $M_{Halo}/M_{baryon}$ and $M_{baryon}/M_*$ if the problems with relative density and assembly time of the halos are ignored.

\subsubsection{Other Explanations}
\label{subsubsec:imf}

Are there other ways to rapidly change the halo mass to UV luminosity ratio in order to explain the observed mismatch?  One possibility is to allow for rapid evolution in the initial mass function (IMF).  The IMF is expected to be substantially the same across cosmic time, even at $Z \sim 0.01$, similar to what is seen in the lowest-metallicity stellar populations in nearby dwarf galaxies \citep{Dias2010,Fagotto1994}.  Thus, a much top-heavier IMF would either require a new understanding of early star formation or even the possibility that at $z = 6$, there remain residual massive stars from early stellar populations that formed at even lower-metallicities.  Because massive main sequence stars have very short lifetimes, this is also unlikely.  

The other possibility is that the halo mass to stellar mass ratio might have evolved.  The standard ratio comes from a combination of expecting that 10\% of baryons have condensed into stars \citep{Leathaud2012} and that there is a 6:1 dark matter to baryonic matter ratio \citep{Planck2015}.  To change this ratio by 0.8 dex would either require a complete absence of dark matter at $z = 8$ or that nearly 100\% of baryons end up in stars at high redshift, both of which would likely require new physics that alters halo mass functions as well.  

In conclusion, a possible solution for the disagreement between hierarchical merging and observation is a change in our theory of early star formation, so that we cannot easily convert between observed luminosities and halo mass.  Such a solution would be intriguing in its own right, with implications discussed further in \S~\ref{sec:discussion}.

\subsubsection{Dangers of Abundance Matching}
\label{subsubsec:abundancematching}

Abundance matching compels agreement between observation and theory, even though it often comes at the expense of having a strong physical motivation for such a model.  As a result, it can be difficult to understand exactly what it means for abundance matching to have succeeded or failed.  Since any pair of continuous functions can be matched in a way that is empirically correct at that redshift, in some sense abundance matching will always be successful, and it is difficult to select only the physically meaningful matches from a technique that includes no underlying physics.

Perhaps, then, abundance matching should really be thought of as an extension of the physical model that produces a halo mass function.  Combining that model with observations produces additional constraints that essentially become part of the theory, and must be predicted for that theory to be complete.  For example, in this work we have shown that $\Lambda$CDM requires rapid, major changes in the properties of star-forming galaxies between $z = 4$ and $z = 8$.  A logical conclusion is that either a mechanism for those changes must be incorporated into $\Lambda$CDM or the model must be rejected.  As shown earlier in this section, producing such a mechanism appears difficult, but there is also a large space of possible models that could be developed.

One of the reasons for that large parameter space is that the galaxy luminosity function is a degenerate combination of many different properties.  This is why we can conclude that $\Lambda$CDM requires a sharp break in typical galactic properties between $z = 4$ and $z = 8$, but cannot specify precisely which properties must change rapidly.  Many parameters, including the initial mass function, stellar population age, star formation efficiency, extinction and differential clustering between baryons and dark matter combine to produce this discontinuous behavior.  As a result, it is possible to pick nearly any subset of these parameters and abundance match in a way that avoids rapid evolution, but always at the cost of sharp changes in some of the others.

For example, \citet{Trac2015} report that the star formation efficiency is sharply variable as a function of both mass and redshift, and \citet{Finkelstein2015} instead express it as an increasing baryon fraction (see also \S~\ref{subsec:asp}).  \citet{Behroozi2015} choose to report an overall halo mass-to-light ratio, allowing that to vary sharply with both mass and redshift.  In an alternative approach, \citet{Mashian2015} match observation with theory in order to make prediction at $z >10$ by removing the continuity requirement, so that galaxies have no consistent history, allowing mass to be both added and subtracted in any quantities necessary to match halo mass functions at each redshift.

In this work, galactic evolution is expressed with respect to redshift, but because observations at different redshifts are also viewing different ranges in both stellar and (presumably) halo mass, we note that in many studies this has instead been expressed with respect to a combination of both mass and redshift.  Similarly, where necessary we compare different quantities by assuming that galactic scaling ratios determined at lower redshift continue to hold, even though sharp changes in those relations might also be consistent with all existing observations.  Ultimately, the proper way to express this effect depends upon its cause, and as we have shown, the most likely astrophysical effects seem incapable of producing such rapid evolution given our current understanding.  

\section{Producing Massive Galaxies in Early Halos}
\label{sec:models}

Having considered the extent to which observations of high-redshift galaxies might allow multiple interpretations, it is important to do the same for our theoretical understanding of hierarchical merging.  Although the baryonic physics involved in star formation is quite complex and there are multiple definitions of halo mass used in describing the results of simulations, there is broad consensus on how dark matter behaves and on the number density and size of the massive halos that they form.  
Explaining an observed number density of galaxies in terms of the density of formed halos depends upon three parameters: (1) The fraction of halos containing a galaxy, or halo occupation rate which is measured to be $\sim$40\% at $z\sim5$ \citep{Hildebrandt2009}; (2) The fraction of baryons converted into stars, which can be parameterized and is 10\% at low redshift \citep{Leathaud2012}; and (3) The amount of time required after virialization for those stars to have formed which will translate into a Halo-mass to light ratio and is parameterized by stellar population models.  We display in Table 1 
various combinations of these parameters could produce the observed number density of $2 \times 10^{-5} \textrm{ Mpc}^{-3}$ for $M_*=10^{11} M_\odot$ galaxies at $z=5.5$.
\begin{table}
\begin{tabular}{c c c c c c}
Halo Occ. & Baryon Frac & SF Time & $M_{halo}/M_\odot$ & $z_{form}$ \\ 
\hline
10\% & 100\% & Instant & $5 \times 10^{11}$ & 5.5 \\
100\% & 30\% & Instant & $2 \times 10^{12}$ & 5.5 \\
100\% & 100\% & 150 Myr & $5 \times 10^{11}$ & 6.4 \\
100\% & 10\% & (-1.1 Gyr) & $5 \times 10^{12}$ & 3.0 \\
\end{tabular}
\label{tab:combos}
\caption{Various combinations of parameters producing the observed number density of $10^{11} M_\odot$ galaxies at $z=5.5$, as described in \S~\ref{sec:models}.}
\end{table}

We find that each of these combinations requires implausible physics, such as 100\% of baryons being turned into stars instantly upon halo virialization or 10\% of baryons forming stars over 1 Gyr before the dark matter halo virializes, or contradicts other observational results.  For example, several combinations require halo masses below $10^{12} M_\odot$.  However, at $3.1 < z < 4.7$, \citet{Hildebrandt2009} used clustering measurements in CFHTLS to find that galaxies at 25.5 mag are found in halos of $\log M_{halo}/M_\odot = 12.3$.  A similar ratio would yield $\log M_{halo}/M_\odot = 12.8$ for massive galaxies at $z = 6$.  \citet{Finkelstein2015} use CANDELS observations to argue that the dark matter to baryon ratio decreases towards higher redshift, so that the halo mass to stellar mass ratio is 50:1 at $z = 6$ and 40:1 at $z = 7$.  This would correspond to $\log M_{halo}/M_\odot = 12.7$.  

In summary, solutions that give a plausible halo occupation fraction require an implausibly short timescale for star formation, a much higher fraction of baryons to be converted into stars than the 10\% in current models, or both.  Any solution with a standard ratio between the halo mass and stellar mass of between 50:1 and 100:1 ($M_{halo} = 5 \times 10^{12} - 10^{13} M_{\odot}$) requires most of the star formation to occur well in advance of initial collapse and virialization.

\subsection{Massive Galaxies in Merger Simulations}

Extensive effort has been put into studying the formation of massive galaxies and their dark matter halos through numerical simulations.  The vast difference in both scale and dominant physical processes between hierarchical merging and star formation means that simulations cannot investigate both processes directly, but rather use semi-analytical prescriptions for connecting the properties of massive galaxies to their halos.  

These prescriptions attempt to model baryonic physics on a very macroscopic level and are drawn from lower-redshift relationships observed between galaxies and their host halos.  However, extrapolating these relationships often leads to unphysical results: the Millennium simulation \citep{Springel2005} can produce $M_* = 10^{11} M_\odot$ galaxies at $z = 6$, but they live in dark matter halos with $M = 10^{11.3} M_\odot$.  This is a very small halo mass to stellar mass ratio on several fronts: Theoretically, it would require the baryons to cluster in advance of much of the dark matter, as well as very nearly all baryons to have ended up in stars by $z = 6$.  Observationally, it would be in conflict with the \citet{Hildebrandt2009} and \citet{Finkelstein2015} halo mass to stellar mass ratios discussed in the previous section.

The Illustris simulation \citep{Illustris,Sparre2015,Wellons2015} picks a set of baryonic relationships that avoids these unphysical extrapolations, resulting in a stellar mass function and luminosity function that look similar to the halo mass function.  As a result, the simulated number densities of massive galaxies are consistent with observation out to $M_* \sim 10^{10} M_\odot$ at $z \sim 4-6$, but very few galaxies are produced with $M_* > 10^{10.5} M_\odot$, disagreeing with the observed number densities at that mass and redshift.

 The conclusion from these simulations is that there is broad consensus on the halo mass function, but considerable freedom in matching those halo masses to galactic properties.  Even given that freedom, matching observations would require that either stellar masses are vastly overestimated or there is a sharp disconnect between the baryons in massive galaxies and their dark matter halos.  Specifically, simulations are consistent with typical galaxies at these redshifts, but cannot produce the earliest, most massive galaxies seen at $z > 4$ with the introduction of reasonable physics.  Rather, then, understanding these galaxies appears to require additional physics not yet included in these simulations, whether baryonic physics relating to star formation or high-energy physics altering the timing of massive halo formation.

\section{Discussion}
\label{sec:discussion}

We have shown that recent observations of high-redshift galaxies are inconsistent with current theoretical models of galactic assembly.  As a general principle, when theory and observation disagree, it is historically best to believe the observational result.  However, in this case the observations also rely on untested theoretical assumptions about stellar evolution.  Thus, something is wrong, but what?  We can divide the possible flaws and explanations into three possible categories:

\begin{enumerate}
\item{{\bf Failed Template Fitting or Redshift Determination}: If the measurements are wrong, it is most likely not in the redshifts (as spectroscopic redshifts exist for many, albeit less massive galaxies at $z > 5$) but rather in the assumption that templates derived from lower-redshift galaxies can be used at $z = 6$.  This implies that the halo mass to monochromatic luminosity ratio changes sharply above $z=4$.  As discussed in \S~\ref{subsec:mlevolve}, the most likely explanation for this evolution would be a sharply top-heavier IMF at higher redshifts.  This can potentially be tested in the near future using supernova rates, and certainly following the launch of JWST.

Later spectroscopy confirmed that low-redshift templates yielded correct results for inferred quantities such as stellar mass out to $z < 3$.  If this breaks down by $z \sim 6$, it would mean that purely photometric surveys are now insufficient, as new models must be developed and observationally tested by JWST.  It is inevitable that at some point astronomers must encounter this problem, but would be unpleasant to discover that it happens at a lower redshift than currently believed.}
\item{{\bf New Clustering Physics}: Another possibility is that halos indeed collapse earlier than allowed by current models, something that would simultaneously solve the high-redshift massive quasar problem as well.  Current collapse times are derived from gravity acting on perturbations that can be confirmed using cosmic microwave background measurements, and therefore a much more rapid collapse of the halos would require the introduction of new high-energy physics.  Possible solutions here might then provide exciting new constraints on the nature of dark energy or dark matter.  It should be noted that many dark matter models under current consideration are warm dark matter, which would suppress the $z \sim 6$ halo mass function rather than enhance it \citep{Gao2007}.  Dark energy with $w > -1$ could enhance early structure formation, although cosmic microwave background observations \citep{Planck2015} create considerable tension with the $w > -0.95$ required to solve this problem \citep{OHara2006,Gladders2007}}
\item{{\bf Early Star Formation}: A baryonic solution is instead to allow main sequence star formation much earlier than the initial collapse of halos.  Such ideas would need to evade difficult constraints from both low-redshift observations, as well as solve the problem of cooling to form small stars at low metallicities.  We further note that if initial stars form in small clumps rather than in fully-formed protogalaxies, the ``clumpyness factor'' in reionization \citep{Ouchi2009} will be much higher than expected, resulting in far more rapid reionization from star formation alone than currently expected.  If so, this will be evident in infrared background fluctuation measurements.}
\end{enumerate}
 
All three answers carry major consequences for both our current understanding of the initial stages of galactic formation and our future plans for studying high-redshift galaxies.  So, better observations are needed.  There is considerable hope that followup observations will help to determine whether the first of these three explanations is the right one.

Rather than speculate as to which explanation is best, we instead stress that the high-mass objects coming out of high-redshift surveys are now critically important.  Future surveys should concentrate on finding and characterizing these objects in sufficient numbers to constrain how these galaxies and their halos co-evolve.  Since the earliest, most massive galaxies are rare, wide area surveys on the $>1$ degree scale will be needed.  These objects are no longer merely an extra point or two in the last panel of a figure, but rather pose a key problem at the heart of high-redshift extragalactic astronomy, and need to be given corresponding attention.

\acknowledgements
The authors would like to thank Michaela Bagley, Peter Behroozi, Rychard Bouwens, Andreas Faisst, Steven Finkelstein, Hendrik Hildebrandt, Alexander Karim, Olivier Le Fevre, Nick Lee, Surhud More, Dave Sanders, Johannes Staguhn, and Dominic Yurk for helpful comments.

\bibliographystyle{apj}

\end{document}